\documentclass[10pt]{article}

\usepackage{amsmath}
\usepackage{amssymb,latexsym}

\title{\large Nonperturbative renomalization group for Einstein gravity with matter}
\author{L. N. Granda\thanks{e-mail: ngranda@univalle.edu.co} \\ {\it Departamento de F\'isica, Universidad del Valle, A.A. 25360},\\ {\it Cali, Colombia}}
 
\begin{document}
\maketitle
\begin{abstract}
we investigate the exact renormalization group (RG) in Einstein gravity coupled 
to N-component scalar field, working in the effective average action formalism 
and background field method. The truncated evolution equation is obtained for the Newtonian 
and cosmological constants. We have shown that screening or antiscreening 
behaviour of the gravitational coupling depends cricially on the choice of 
scalar-gravitational $\xi $ and the number of scalar fields.\vspace{1cm}
\end{abstract}


There has recently been much activity in the study of nonperturvative RG dynamics in field 
theory models ( for recent references, see \cite{Bonini} ). One of the versions of 
nonpertubative RG based on the effective average action has been developed 
in ref.\cite{Reuter} for Einstein gravity  ( the gauge dependence problem in 
this formalism has been studied in ref. \cite{Falkenberg} ) and in \cite{Granda} for gauged 
supergravity. The nonperturbative RG equation for cosmological and Newtonian
coupling constants have been obtained in ref.\cite{Reuter}  for the Einstein gravity and 
in ref \cite{Bytsenko} for $R^2$ gravity. The comparison between quantum correction to 
Newtonian coupling from nonperturvative RG \cite{Reuter} and from effective field theory 
technique \cite{Donoghue} has been done.

It is quite interesting to study nonperturbative RG ( or evolution equation ) for the gravity 
coupled to N-component scalar field in orden to evaliate the influence of the scalar coupling 
and the number of scalar in the nonperturbative behaviour of the Newtonian and cosmological 
constants. We follow the formalism of ref. \cite{Reuter} developed for gravitational theories. 
The basic elements of it are the background field method ( see \cite{Buchbinder} for a review ) 
and the truncated nonperturbative evolution equation for the effective average action 
\cite{Reuter}. We will start from the theory with the following action.

\begin {equation}
S=\int d^{4}x \sqrt{-g}\Bigr[ \frac{1}{16\pi \bar G}\bigr(-R+2\lambda  \bigl)+\frac{1}{2}\partial_{\mu }\phi ^{i}\partial^{\mu }\phi _{i}+\frac{1}{2}\xi R\phi ^{i}\phi _{i}  \Bigl],
\end{equation}
where $i=,...,N$ and nonperturbative Einstein gravity is considered to be valid below some UV 
scale $\Lambda $. We will use the truncation \cite{Reuter}, \cite{Falkenberg}
\begin{equation}
\bar G\rightarrow G_{k}=\frac{\bar G}{Z_{Nk}},           \lambda\rightarrow\lambda_k
\end{equation}

Following the approach of ref.\cite{Reuter} we will write the evolution equation 
for the effective average action $\Gamma _{k}[g,\bar g]$ defined at non zero momentum ultraviolet $k$ below some cut-off $\Lambda_{cut-off}$.\\
The truncated form of such evolution equation is
\begin{equation}\label{gran2}
\begin{aligned}
\partial_{t}\Gamma _{k}[g,\bar g]=&\frac{1}{2}Tr[(\Gamma^{(2)} _{k}[g,\bar g]+R^{grav}_{k}[\bar g])^{-1}\partial_{t}R^{grav}_{k}[\bar g]\bigl] \\ &-Tr[(-M[g,\bar g]+R^{gh}_{k}[\bar g])\partial_{t} R^{gh}_{k}[\bar g]],
\end{aligned}
\end{equation}

where $t=Ln k, k$ is the nonzero momentum scale, $R_{k}$ are cutt-off, $M[g,\bar g]$ are ghost o
perators, $\bar g_{\mu \nu}$ is the background metric, $g_{\mu \nu }=\bar g_{\mu \nu }+
h_{\mu \nu }$, where $h_{\mu \nu }$ is the quantum field. $\Gamma ^{(2)}_{k}$ is the Hessian 
of $\Gamma _{k}[g,\bar g]$ with respect to $g_{\mu \nu }$ at fixed $\bar g_{\mu \nu }$ (for 
more details, see\cite{Reuter}).\\
Projecting the evolution equation on the space with low-derivatives terms, one gets the 
left-hand side of the truncated evolution equation (\ref {gran2}) as follows:
\begin{equation}\label{gran3}
\partial_{t}\Gamma _{k}[g,\bar g]=2k^{2}\int d^{4}x\sqrt{g}[-R(g)\partial_{t}Z_{Nk}+
2\partial_{t}(Z_{Nk}\lambda _{k})],
\end{equation}
with $k^{2}=\frac{1}{32\pi \bar G}$. The initial conditions for $Z_{Nk}$, $\lambda _{k}$ are 
chosen in  the same way as in $\cite{Reuter}$.\\
In the right-hand side of the evolution equation $(2)$ we need the second functional derivate 
of $\Gamma _{k}[g,\bar g]$ at fixed background $\bar g_{\mu \nu }$,
\begin{equation}\label{gran4}
\begin{aligned}
\Gamma ^{(2)}_{k}[g, g]&=Z_{Nk}k^{2}\int d^4x\sqrt{g} \Bigl\{\frac{1}{2}\hat h_{\mu \nu }
\bigr[-\Box-2\bar \lambda _{k}+R\bigr]\hat h_{\mu \nu }-
\frac{1}{8}\phi [-\Box-2\bar \lambda _{k}]\phi\\ & 
+\frac{1}{2}{\tilde \phi}_{i}[-\Box+\xi R] \tilde \phi^{i}-R_{\mu \nu }h^{\nu \rho }h^{\mu }_{\rho }
+R_{\alpha \beta \mu \nu }h^{\beta \nu }h^{\alpha \mu }\Bigl\}
\end{aligned}
\end{equation}
where we have rescaled the scalar field
\begin{equation}
\phi _{i}\rightarrow \tilde \phi _{i}=32\pi \bar G\phi _{i}.
\end{equation}
In orden to calculate the effective average action, we have to specify some curve background.\\
A useful choice is a maximal symmetric space where the curvature $R$ is the external parameter characterizing the space. In such space we have
\begin{equation}\label{gran5}
\begin{aligned}
\Gamma ^{(2)}_{k}[g, g]&=Z_{Nk}k^{2}\int d^4x\sqrt{g} 
\Bigl\{\frac{1}{2}\hat h_{\mu \nu }\bigr[-\Box-2\bar \lambda _{k}+
\frac{2}{3}R\bigr]\hat h_{\mu \nu }\\ &-\frac{1}{8}\phi [-\Box-2\bar \lambda _{k}]\phi +
\frac{1}{2}{\tilde \phi}_{i}[-\Box+\xi R] \tilde \phi^{i}\Bigl\}
\end{aligned}
\end{equation}
The complete average effective action wiht $\bar g=g$ containing gravitational, ghosts and 
scalar contributions is calculated with final result
\begin{equation}\label{gran6}
\begin{aligned}
\Bar \Gamma_{k}=\frac{1}{2}& Tr_{T}Ln\Biggl[Z_{Nk}\Bigl(-\Box -2 \lambda _{k}+
\frac{2}{3}R+k^{2}R^{(0)}(-\Box/k^{2})\Bigr)\Biggr]\\ &+
\frac{1}{2}Tr_{S}Ln\Bigl[Z_{Nk}(-\Box-2\lambda _{k}+
k^{2}R^{0}(-\Box/k^{2}))]\\ & -Tr_{V}Ln[(-\Box-\frac{1}{4}R+k^{2}R^{(0)}(-\Box/k^{2}))]\\ &+
\frac{N}{2}Tr_{S}Ln[Z_{Nk}(-\Box+\xi R+k^{2}R^{(0)}(-\Box/k^{2}))]
\end{aligned}
\end{equation}
Now we want to find the RHS of the evolution equation . To this end, we differentiate the 
average action (6) with respect to $t$. Then we expand the operators in (6) with respect to 
the curvature $R$ because we are only interested in terms of orden $\int d^{4}x\sqrt{g}$ and 
$\int d^{4}x\sqrt{g}R$:
\begin{equation}
\begin{aligned}
\partial_{t}\Gamma _{k}[g,g]=& Tr_{T}\biggl[{\cal N}\Bigl( A+\frac{2}{3}R\Bigr)^{-1} \Biggr]+
Tr_{S}[{\cal N}A^{-1}]\\ & -2Tr_{V}\Biggl[ {\cal N}_{0}\Bigl( {\cal A}_{0}-\frac{1}{4}R    
 \Bigr)^{-1} \Biggr]+N Tr_{S}[{\cal N}_{0}({\cal A}_{0}+\xi R)^{-1}],
\end{aligned}
\end{equation}
where\\
\begin{equation}
{\cal N}=\frac{\partial_{t}[Z_{Nk}k^{2}R^{(0)}(z)]}{Z_{Nk}},
\end{equation}

\begin{equation}
{\cal A}=-\Box +k^{2}R^{(0)}(z)]-2\lambda _{k}.
\end{equation}

The operators ${\cal N}_{0}$ and ${\cal A}_{0}$ are defined from (8) with $\lambda _{k}=0$ 
and $Z_{Nk}=1$ (see \cite{Reuter}, \cite{Falkenberg}). Here the variable $z$ replaces 
$-\Box/k^{2}$ and $\eta _{N}(k)=-\partial_{t}(Ln Z_{Nk})$. Note that as a cut-off we use 
the same function as in ref.\cite{Reuter}: $R^{(0)}(z)=\frac{z}{exp[z]-1}$. The above steps 
lead then to
\begin{equation}
\begin{aligned}
\partial_{t}\Gamma _{k}[g,g]=& Tr_{T}\biggl[{\cal N}A^{-1} \Biggr]+Tr_{S}[{\cal N}A^{-1}] 
-2Tr_{V}\Biggl[ {\cal N}_{0}{\cal A}_{0}^{-1} \Biggr]\\ &+N Tr_{S}[{\cal N}_{0}{\cal A}^{-1}_{0}]-R \Bigl\{ \frac{2}{3}Tr_{T}[{\cal N}{\cal A}^{-1}]+\frac{1}{2}Tr_{V}[{\cal N}_{0}{\cal A}_{0}^{-2}]\\ &+\xi N Tr_{S}[{\cal N}_{0}{\cal A}_{0}^{-1}]\Bigr\}+O(R^{2}),
\end{aligned}
\end{equation}
As a next step we evaluate the traces. We use the heat kernel expansion which for an arbitrary 
function of the covariant Laplacian $W(D^{2})$ reads
\begin{equation}
\begin{aligned}
Tr_{j}[W(-D^{2})]=& (4\pi )^{-1}tr_{j}(I)\Biggl\{ Q_{2}[W]\int d^{4}x\sqrt{g}\\ & + 
\frac{1}{6}Q_{1}[W]\int d^{4}x\sqrt{g}R+O(R^{2})\Biggr \},
\end{aligned}
\end{equation}
where by $I$ we denote the unit matrix in the space of field on which $D^{2}$ acts. Therefore $tr_{j}(I)$ simply counts the number of independent degrees freedom of the field. The sort $j$ of fields enters $(15)$ via $tr_{j}(I)$ only. Therefore, we will drop the index $j$ after the evaluation of the traces in the heat kernel expansion.\\
The functionals $Q_{n}$ are the Mellin transforms of W, 
\begin{equation}
\begin{aligned}
Q_{n}[W]=\frac{1}{\Gamma _{(n)}}\int^{\infty}_{0} dzz^{n-1}W(z), \hspace{1cm}(n > 0)
\end{aligned}
\end{equation}
Now we have it perform the heat kernel expansion (10) in eq.(9). This leads to a polynomial 
in $R$ which is the RHS of the evolution equation (2).\\
By the comparison of coefficients with the LHS of the evolution equation (3), we obtain  
at the orden $\int d^{4}x\sqrt{g}$
\begin{equation}
\begin{aligned}
\partial_{t}(Z_{Nk}\bar  \lambda _{k})=\frac{1}{4k^{2}}\frac{1}{(4\pi )^{2}}\{ 10Q_{2}[{\cal N}/{\cal A}]-
8Q_{2}[{\cal N}_{0}/{\cal A}_{0}]+NQ_{2}[{\cal N}_{0}/{\cal A}_{0}]\}
\end{aligned}
\end{equation}
and at the orden $\int d^{4}x\sqrt{g}R.$
\begin{equation}
\begin{aligned}
\partial_{t}Z_{Nk}&=-\frac{1}{12k^{2}}\frac{1}{(4\pi )^{2}}\{10Q_{1}[{\cal N}/{\cal A}]-
8Q_{1}[{\cal N}_{0}/{\cal A}_{0}]\\ &+NQ_{1}[{\cal N}_{0}/{\cal A}_{0}]-
36Q_{2}[{\cal N}/{\cal A}^{2}]-12Q_{2}[{\cal N}_{0}/{\cal A}^{2}_{0}]-
6\xi NQ_{2}[{\cal N}_{0}/{\cal A}^{2}_{0}]\}.
\end{aligned}
\end{equation}

The cutt-off-dependent integrals are defined in \cite{Granda}

\begin{equation}
\varPhi ^{P}_{n}(w)=\frac{1}{\Gamma (n)}\int ^{\infty }_{0}dzz^{n-1}
\frac{R^{(0)}(z)-zR^{(0)'}(z)}{[z+R^{(0)}(z)+w]^{P}},
\end{equation}

\begin{equation}
\tilde \varPhi ^{P}_{n}(w)=\frac{1}{\Gamma (n)}\int ^{\infty }_{0}dzz^{n-1}
\frac{R^{(0)}(z)}{[z+R^{(0)}(z)+w]^{P}},
\end{equation}
for $n > 0$. It follows that $\varPhi^{P}_{0}(w)=\tilde \varPhi^{P}_{0}(w)=(1+w)^{-P} $ 
for $n=0$. We can rewrite eqs. (12) and (13) in terms of $\varPhi $ and $\tilde  \varPhi $. 
This leads to the following system of equations:
\begin{equation}
\begin{aligned}
\partial_{t}(Z_{Nk}\bar \lambda _{k})&
=\frac{1}{4k^{2}}\frac{1}{(4\pi )^{2}}k^4\{10\varphi^{1}_{2}(-2\bar \lambda _{k}/k^{2})-
8\varPhi^{1}_{2}(0)\\ & +N\varPhi^{1}_{2}(0)-
5\eta _{N}(k)\tilde \varPhi^{1}_{2}(-2\bar \lambda _{k}/k^{2})\},
\end{aligned}
\end{equation}

\begin{equation}
\begin{aligned}
\partial_{t}Z_{Nk}&=-\frac{1}{12k^{2}}
\frac{1}{(4\pi )^{2}}k^{2}\{10\varPhi^{1}_{1}(-2\bar \lambda _{k}/k^{2})+
(N-8)\varPhi^{1}_{1}(0)\\ &-36\varPhi^{2}_{2}(-2\bar \lambda _{k}/k^{2})-
(12+6\xi N)\varPhi^{2}_{2}(0)\\ &-5\eta _{N}(k)\tilde \varPhi^{1}_{1}(-2\bar \lambda _{k}/k^{2})+18\eta _{N}(k)\tilde \varPhi^{2}_{2}(-2\bar \lambda _{k}/k^{2})\} .
\end{aligned}
\end{equation}

Now we introduce the dimensionless, renormalized Newtonian constant and cosmological constant

\begin{equation}
g_{k}=k^{2}G_{k}=k^{2}Z^{-1}_{Nk}\bar G,\hspace{1cm}\lambda _{k}=k^{-2}\bar \lambda _{k}.
\end{equation}
Here $G_{k}$ is the renormalized Newtonian constant at scale $k$. The evolution equation 
for $g_{k}$ reads then
\begin{equation}
\partial_{t} g_{k}=[2+\eta _{N}(k)]g_{k}.
\end{equation}
from $(16)$ we find the anomalous dimension $\eta _{N}(k)$
\begin{equation}
\eta _{N}(k)=g_{k}B_{1}(\lambda _{k})+\eta _{N}(k)g_{k}B_{2}(\lambda _{k})
\end{equation}
where\\
\begin{equation}
\left\{\begin{aligned}&B_{1}(\lambda _{k})=\frac{1}{6\pi }[10\varPhi^{1}_{1}(-2\lambda _{k})+
(N-8)\varPhi^{1}_{1}(0)-36\varPhi^{2}_{2}(-2\lambda _{k})-(12+6\xi N)\varPhi^{2}_{2}(0)],\\ 
&B_{2}(\lambda _{k})=\frac{1}{6\pi }[18\tilde \varPhi^{2}_{2}(-2\lambda _{k})-
5\tilde \varPhi^{1}_{1}(-2\lambda _{k})].
\end{aligned}\right.
\end{equation}\\
Solving (19)\\
\begin{equation}
\eta _{N}(k)=\frac{g_{k}B_{1}(\lambda _{k})}{1-g_{k}B_{2}(\lambda _{k})}
\end{equation}

we see that the anomalous dimension $\eta _{N}$ is a nonperturbative quantity. From $(15)$ we obtain the evolution equation for the cosmological constant
\begin{equation}
\begin{aligned}
\partial _{t}(\lambda _{k})=&-[2-\eta _{N}(k)]\lambda _{k}+\frac{1}{2\pi }g_{k}[10\varPhi^{1}_{2}(-2\lambda _{k})+\\ &+(N-8)\varPhi^{1}_{2}(0)-5\eta _{N}(k)\tilde \varPhi^{1}_{2}(-2\lambda _{k})].
\end{aligned}
\end{equation}
Equations $(18)$ and $(22)$ together with $(20)$ give the system of differential aquations for two $k$-dependent constants $\lambda _{k}$ and $g_{k}$. These equations determine the value of the running Newtonian constant and cosmological constant at the scale $k\ll \Lambda _{cut-off}.$ The  above evolution equations include nonperturbative effects which go beyond  a simple one-loop calculation.\\
Next, we estimate the qualitative behaviour of the running Newtonian constant as the above system of RG equations is too complicated and cannot be solve analytically. To this end we assume that the cosmological constant is much smaller than the IR cut-off scale, $\lambda _{k}\ll k^{2}$, so we can put $\lambda _{k}=0$ that simplifies eqs. $(18)$ and $(20)$. After that, we make an expansion in powers of $(\bar Gk^{2})^{-1}$ keeping oly the first term (\textit{i,e.} we evaluate the functions $\varPhi^{P}_{n}(0)$) and $\tilde \varPhi^{P}_{n}(0)$) and finally obtain (with $g_{k}\sim k^{2}\bar G$)
\begin{equation}
\begin{aligned}
G_{k}=G_{0}[1-w\bar Gk^{2}+...]
\end{aligned}
\end{equation}
where \\
\begin{equation}
\begin{aligned}
w&= \frac{1}{12\pi }[(48+6\xi N)\varPhi^{2}_{2}(0)-(2+N)\varPhi^{1}_{1}(0)]\\ &=\frac{1}{12\pi }\Bigl[ (48+6\xi N)-(2+N)\frac{\pi ^{2}}{6}\Bigr],
\end{aligned}
\end{equation}\\
where $\varPhi^{1}_{1}(0)=\pi ^{2}/6$ and $\varPhi^{2}_{2}(0)=1$.\\
In the case of Einstein gravity, a similar solution has been obtained in refs. \cite{Reuter}, \cite{Falkenberg}. From $(19)$ we can write the inequality
\begin{equation}
\begin{aligned}
\Bigl[\frac{4}{\pi } -\frac{\pi }{36} \Bigr]+N \Bigl[\frac{\xi }{2\pi }-\frac{\pi }{72} \Bigr]>0,
\end{aligned}
\end{equation}\\
which means that the limit $N\rightarrow \infty $ and for values of $\xi >\pi^{2}/36$, the coefficient $w>0$, which indicates that the Newtonian coupling decreases as $k^{2}$ increases; \textit{i,e.} we find that the gravitational coupling is antiscreening. For $\xi <\pi ^{2}/36$ we have $w<0$ and a screening effect follows. For small $N$ the dependence of the sign of $w$ on $\xi $ is given by eq. $(24)$. Hence, we proved that screening or antiscreening behaviour of the gravitational constant depends crucially on the choice of $\xi $ (which may be dictated by asymptotic freedom \cite{Buchbinder}) or the number of fields. That may lead to interesting coslogical consequences in the early universe.\\\\
\centerline{***}\\\\
This work was supported by \textsl{COLCIENCIAS} (Colombia) Project N0.1106-05-393-95.\\
I thank S. D. Odintsov for stimulating discussions.\\

\end{document}